\documentclass[reprint, 
               superscriptaddress, 
               amsmath, 
               amssymb, 
               aps, 
               floatfix]{revtex4-2}

\usepackage[english]{babel}
\usepackage[utf8]{inputenc}
\usepackage[dvipsnames]{xcolor}
\usepackage{placeins}

\usepackage{makecell,
            graphicx,
            amsfonts, 
            tikz, 
            geometry, 
            import, 
            hyperref, 
            amsthm, 
            appendix, 
            physics}

\makeatletter 
\renewcommand{\fnum@figure}{\textbf{Figure~\thefigure}}
\makeatother

\usepackage{algorithm2e}

\SetAlgoCaptionSeparator{.}
\SetAlgoCaptionLayout{raggedright}

\usetikzlibrary{quantikz2,
                decorations.pathmorphing, 
                decorations.markings,
                shapes, 
                external,
                graphs,
                positioning,
                calc}

\pgfdeclarelayer{bg} 
\pgfdeclarelayer{fg} 
\pgfsetlayers{bg, main, fg}  

\def\rotateclockwise#1{
  \newdimen\xrw
  \pgfextractx{\xrw}{#1}
  \newdimen\yrw
  \pgfextracty{\yrw}{#1}
  \pgfpoint{\yrw}{-\xrw}
}

\def\rotatecounterclockwise#1{
  \newdimen\xrcw
  \pgfextractx{\xrcw}{#1}
  \newdimen\yrcw
  \pgfextracty{\yrcw}{#1}
  \pgfpoint{-\yrcw}{\xrcw}
}

\def\outsidespacerpgfclockwise#1#2#3{
  \pgfpointscale{#3}{
    \rotateclockwise{
      \pgfpointnormalised{
        \pgfpointdiff{#1}{#2}}}}
}

\def\outsidespacerpgfcounterclockwise#1#2#3{
  \pgfpointscale{#3}{
    \rotatecounterclockwise{
      \pgfpointnormalised{
        \pgfpointdiff{#1}{#2}}}}
}

\def\outsidepgfclockwise#1#2#3{
  \pgfpointadd{#2}{\outsidespacerpgfclockwise{#1}{#2}{#3}}
}

\def\outsidepgfcounterclockwise#1#2#3{
  \pgfpointadd{#2}{\outsidespacerpgfcounterclockwise{#1}{#2}{#3}}
}

\def\outside#1#2#3{
  ($ (#2) ! #3 ! -90 : (#1) $)
}

\def\cornerpgf#1#2#3#4{
  \pgfextra{
    \pgfmathanglebetweenpoints{#2}{\outsidepgfcounterclockwise{#1}{#2}{#4}}
    \let\anglea\pgfmathresult
    \let\startangle\pgfmathresult

    \pgfmathanglebetweenpoints{#2}{\outsidepgfclockwise{#3}{#2}{#4}}
    \pgfmathparse{\pgfmathresult - \anglea}
    \pgfmathroundto{\pgfmathresult}
    \let\arcangle\pgfmathresult
    \ifthenelse{180=\arcangle \or 180<\arcangle}{
      \pgfmathparse{-360 + \arcangle}}{
      \pgfmathparse{\arcangle}}
    \let\deltaangle\pgfmathresult

    \newdimen\x
    \pgfextractx{\x}{\outsidepgfcounterclockwise{#1}{#2}{#4}}
    \newdimen\y
    \pgfextracty{\y}{\outsidepgfcounterclockwise{#1}{#2}{#4}}
  }
  -- (\x,\y) arc [start angle=\startangle, delta angle=\deltaangle, radius=#4]
}

\def\corner#1#2#3#4{
  \cornerpgf{\pgfpointanchor{#1}{center}}{\pgfpointanchor{#2}{center}}{\pgfpointanchor{#3}{center}}{#4}
}

\def\hedgeiii#1#2#3#4{
  \outside{#1}{#2}{#4} \corner{#1}{#2}{#3}{#4} \corner{#2}{#3}{#1}{#4} \corner{#3}{#1}{#2}{#4} -- cycle
}

\def\hedgem#1#2#3#4{
  
  \outside{#1}{#2}{#4}
  \pgfextra{
    \def\hgnodea{#1}
    \def\hgnodeb{#2}
  }
  foreach \c in {#3} {
    \corner{\hgnodea}{\hgnodeb}{\c}{#4}
    \pgfextra{
      \global\let\hgnodea\hgnodeb
      \global\let\hgnodeb\c
    }
  }
  \corner{\hgnodea}{\hgnodeb}{#1}{#4}
  \corner{\hgnodeb}{#1}{#2}{#4}
  -- cycle
}

\def\hedgeii#1#2#3{
  \hedgem{#1}{#2}{}{#3}
}

\def\hedgei#1#2{
  (#1) circle [radius = #2]
}

\hypersetup{
    colorlinks=true,
    linkcolor=blue,
    filecolor=magenta,      
    urlcolor=cyan,
    pdfpagemode=FullScreen,
}

\newtheorem{proposition}{Proposition}

\theoremstyle{definition}
\newtheorem{definition}{Definition}

\usetikzlibrary{quantikz2,
                decorations.pathmorphing, 
                shapes, 
                external,
                graphs,
                positioning,
                calc}

\usetikzlibrary{quotes,arrows.meta}
\tikzset{
  annotated cuboid/.pic={
    \tikzset{%
      every edge quotes/.append style={midway, auto},
      /cuboid/.cd,
      #1
    }
    \draw [every edge/.append style={pic actions, densely dashed, opacity=0}, pic actions]
    (0,0,0) coordinate (o) -- ++(-\cubescale*\cubex,0,0) coordinate (a) -- ++(0,-\cubescale*\cubey,0) coordinate (b) edge coordinate [pos=1] (g) ++(0,0,-\cubescale*\cubez)  -- ++(\cubescale*\cubex,0,0) coordinate (c) -- cycle
    (o) -- ++(0,0,-\cubescale*\cubez) coordinate (d) -- ++(0,-\cubescale*\cubey,0) coordinate (e) edge (g) -- (c) -- cycle
    (o) -- (a) -- ++(0,0,-\cubescale*\cubez) coordinate (f) edge (g) -- (d) -- cycle;
    ;
  },
  /cuboid/.search also={/tikz},
  /cuboid/.cd,
  width/.store in=\cubex,
  height/.store in=\cubey,
  depth/.store in=\cubez,
  units/.store in=\cubeunits,
  scale/.store in=\cubescale,
  width=10,
  height=10,
  depth=10,
  units=cm,
  scale=.1,
}

\begin{document}

\RestyleAlgo{boxruled}

\title{Efficient Gate Reordering for Distributed Quantum Compiling in  Data Centers} 

\author{Riccardo Mengoni}
\author{Walter Nadalin}
\author{Mathys Rennela}
\author{Jimmy Rotureau}
\author{Tom Darras}
\affiliation{Welinq, 14 rue Jean Macé, 75011 Paris, France}
\author{Julien Laurat}
\affiliation{Laboratoire Kastler Brossel, Sorbonne Université, CNRS,
ENS-Universite PSL, Collège de France, 4 Place Jussieu, 75005 Paris, France}
\author{Eleni Diamanti}
\affiliation{LIP6, Sorbonne Université, CNRS, 4 Place Jussieu, 75005 Paris, France}
\author{Ioannis Lavdas}
\affiliation{Welinq, 14 rue Jean Macé, 75011 Paris, France}

\date{\today}

\begin{abstract}

Just as classical computing relies on distributed systems, the quantum computing era requires new kinds of infrastructure and software tools.
Quantum networks will become the backbone of hybrid, quantum-augmented data centers, in which quantum algorithms are distributed over a local network of quantum processing units (QPUs) interconnected via shared entanglement. 
In this context, 
it is crucial to develop methods and software that minimize the number
of inter-QPU communications.
Here we describe key features of the quantum compiler \textit{araQne}, which is designed to minimize distribution cost, measured by the number of entangled pairs required to distribute a monolithic quantum circuit using gate teleportation protocols.
We establish the crucial role played by {circuit} reordering strategies, which strongly reduce the distribution cost
compared to a baseline approach.
\end{abstract}

\maketitle

\section{Introduction}

Recent progress in fundamental research, hardware engineering and quantum algorithm development are bringing us closer to practical quantum computing. However, solving real-size problems and relevant industrial use cases requires quantum processing units (QPUs) with a large number of error corrected qubits~\cite{ORUS2019100028,dalzell2023quantumalgorithmssurveyapplications, 
NunezMerino2024,Leclerc_2025, cazals2025identifyinghardnativeinstances}. {This remains} a core challenge for hardware engineers~\cite{mohseni2025buildquantumsupercomputerscaling, sinclair2024faulttolerantopticalinterconnectsneutralatom}, while quantum algorithm researchers are developing methods allowing large scale quantum computation with limited resources~\cite{BARRAL2025100747}.

The concept of Distributed Quantum Computing (DQC)  has emerged as a viable approach to scaling up the computational power of quantum hardware~\cite{Caleffi_2024,PhysRevA.108.032610,Main2025}. To realize this vision, there is a critical need for the development of dedicated software tools~\cite{michele-dqc} and hardware infrastructures~\cite{weaver2024integrated}. Quantum-augmented data centers will implement DQC by interconnecting quantum devices on a local network, in an environment tailored for quantum computation~\cite{shapourian2025quantumdatacenterinfrastructures}. This additionally opens the door to heterogeneous DQC, which aims to harness the strengths of different QPU architectures by interconnecting processors based on different qubit modalities.

\begin{figure*}[t!]
\includegraphics[scale=0.2]{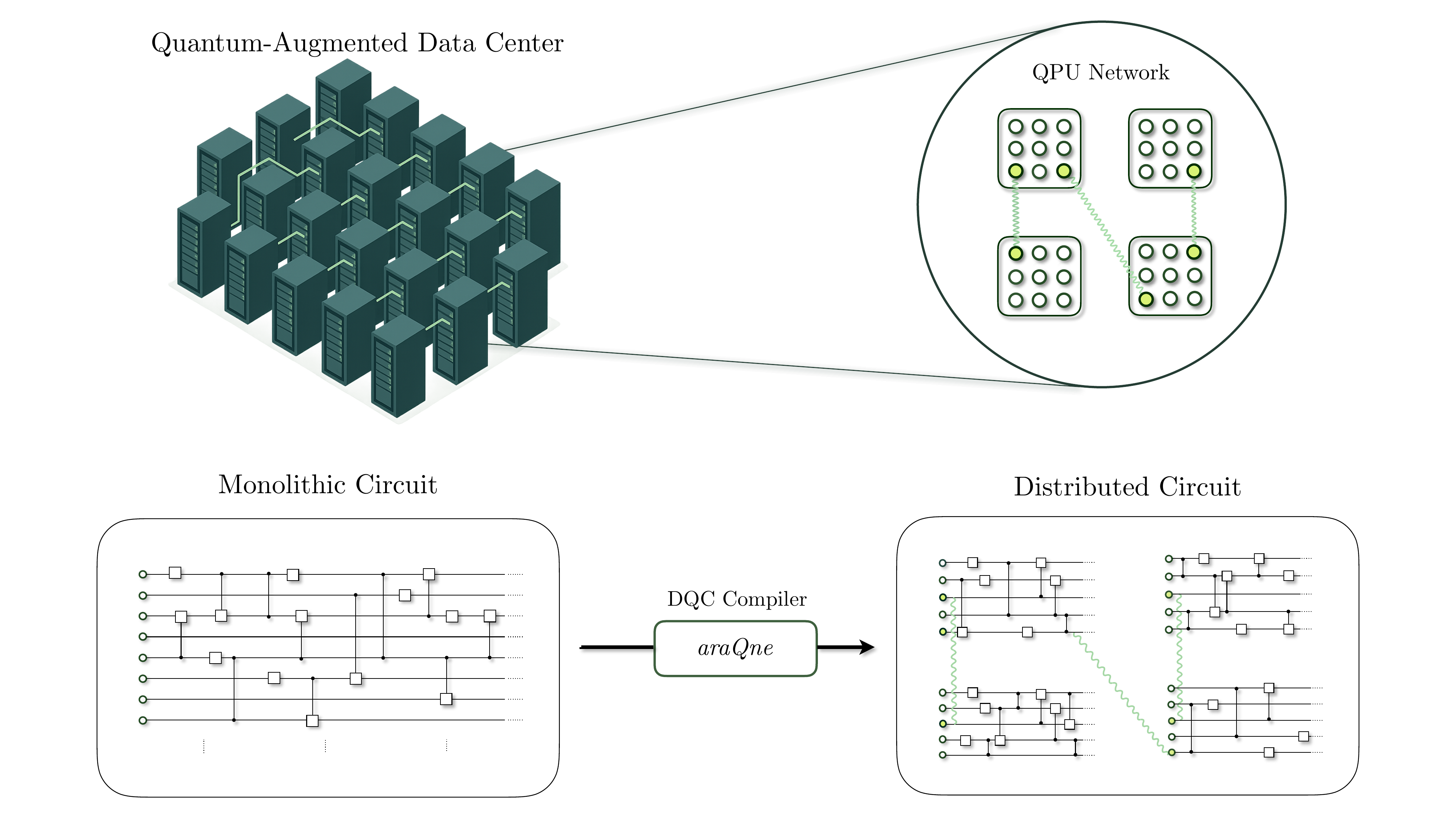}
    \caption{\small Distributed quantum computing as the core of quantum-augmented data centers. The upper part shows the network configuration of QPUs interconnected by entangled states (wavy lines).
    The entangled states are stored in communication qubits (green dots) while data qubits (white dots) are dedicated to processing. The lower part shows the workflow of the quantum compiler that implements the mapping of a monolithic algorithm to such a distributed quantum architecture. The input monolithic algorithm is optimally partitioned with respect to entanglement resources required for the interconnection, which results to its distributed version. 
    }
    \label{DQC}
\end{figure*}

At the core of the DQC approach lies a modular quantum architecture: multiple QPUs are interconnected through both classical and quantum channels, forming a network that functions as a single larger capacity  quantum computer. We consider here that  interconnection between QPUs is achieved through shared entanglement, as sketched in Figure~\ref{DQC}: a given monolithic quantum circuit is partitioned into a set of circuits  executed across multiple QPUs. Within each QPU, qubits are distinguished between data qubits, dedicated to processing, and communication qubits, which store shared entangled states. 

Such entangled states are expected to be scarce, as generating them on demand with high fidelity remains one of the  key challenges in quantum hardware engineering~\cite{over}. For this reason, {a central goal in mapping a quantum algorithm to a distributed architecture is to minimize the amount of the required inter-device entanglement}. To support this objective, software tools,  hereafter called compilers, can be developed to automate the translation of a  quantum algorithm into an equivalent version that can run on a distributed architecture.

Several approaches to distributed compilation have been recently proposed~\cite{martinez,Burt2,Wu_2023,tomesh2023divideconquercombinatorialoptimization,10214316,kaur2025optimizedquantumcircuitpartitioning,wu2022autocommframeworkenablingefficient,PhysRevA.100.032308,Burt}.
Contributions include techniques where qubit assignment is determined by solving a graph partitioning problem that minimizes cross-partition edge weights. For example the approach in \cite{tomesh2023divideconquercombinatorialoptimization} combines graph partitioning with quantum circuit cutting, while the ones in \cite{10214316} and \cite{kaur2025optimizedquantumcircuitpartitioning} process the circuit sequentially searching for feasible qubit and gate teleportation operations. 
Additional work \cite{wu2022autocommframeworkenablingefficient} optimized quantum communication by introducing a gate reordering procedure executed exclusively after the qubit allocation mapping and partitioning phases.
In \cite{martinez, PhysRevA.100.032308}, the authors first introduce the concept of distributable gate packets, which enables the circuit to be represented as a hypergraph. Then following the hypergraph partitioning,  packets are subsequently refined to minimize the  distribution resources. However, such method is constrained to the gate set (\(H\), \(R_z\), and \(CR_z\)) with the corresponding transpilation resulting to additional complexity.
Finally, \cite{Burt, Burt2} further extends the hypergraph-based formulation by showing how to jointly optimize the cost of gate and state teleportation to reduce overall entanglement.

Our work leverages the notion of gate packets \cite{martinez, PhysRevA.100.032308, Burt, Burt2}, which, combined with an efficient circuit optimization method, yields a unique framework of distributed quantum compilation. A characterization of gate packets over a  register of $d$ qubits is provided in terms of  a sequence involving $(d-1)$ two-qubit unitaries, prior to any qubit allocation and circuit partitioning. The novel preprocessing strategy employed in this work reorders the gates of an input circuit, optimizing their grouping into gate packets. The proposed method exploits unitary operator commutation and control-symmetry of logical operations, as well as gate-packet merging, increasing the size of gate packets and thereby reducing the quantum communication overhead required for non-local operations. The above approach is shown to be effectively leveraged for distributed hypergraph-based quantum compilation. The reordering procedure is designed to be agnostic to the specific gate set used to represent the circuit, and hence avoiding complexity introduced by transpilation while enhancing its adaptability across diverse quantum hardware platforms. As a result, our framework provides a flexible and scalable solution suited for the demands of future large-scale quantum computing architectures.


The paper is organized as follows. Section~\ref{sec:workflow} introduces our compiler's workflow. 
We develop a strategy for quantum circuit distribution, which places an emphasis on efficient gate reordering  to lower the distribution cost compared to a baseline approach. We rely on a hypergraph-based framework to map monolithic circuits onto distributed circuits. This strategy has been implemented in the workflow of \textit{araQne}, the compiler for distributed quantum computing developed by Welinq. In Section~\ref{sec:benchmarks} we present results for the distribution cost of several classes of quantum circuits including circuits from the 
widely used QASMBench  1.4 benchmarking suite \cite{qasm}. This comparison establishes the importance of efficient circuit reordering strategies for distributed quantum  compiling. Detailed proofs of the methods  are provided in  Appendix~\ref{proofs}.

\section{Compiler Workflow}
\label{sec:workflow}
In the context of quantum computing, the term compilation refers to strategies that aim to simplify quantum circuits and facilitate their executions, by rewriting them into equivalent circuits with more desirable properties, such as consuming less computational resources, or being executable on specific quantum architectures.

We are interested here in the task of compiling quantum circuits for distributed quantum computers. 
The full workflow corresponding to those compilation routines is called a distributed quantum compiler. 

In this workflow the following tasks are performed:

\begin{enumerate}
    \item \textit{Qubit allocation}: assigning qubits of the input monolithic circuit to the interconnected QPUs.
    \item \textit{Non-local gate scheduling}: implementing non-local gates in a way that minimizes the number of inter-QPU operations. 
\end{enumerate}

The above tasks are typically addressed through \textit{circuit partitioning}, wherein a monolithic quantum circuit is decomposed into subcircuits that can be executed across a network of interconnected QPUs. An additional and crucial feature of our approach is a  \textit{circuit optimization} stage involving a set of routines aimed at minimizing the {quantum resources} needed for distribution.

This section presents the compilation strategies that we have implemented in our compiler's workflow. Here, the computational resource to be optimized is the {\textit{distribution cost}}, that is, the number of EPR pairs consumed in distributing a monolithic quantum circuit. We have developed a greedy {heuristic} algorithm for gate packing: at each stage, the  algorithm aims to minimize the distribution cost of the processed sub-circuit. {The types of  choices made by the greedy algorithm are called \textit{choice properties}. We formalize the choice properties incorporated in our algorithm and provide a justification for their presence.}

\begin{figure}
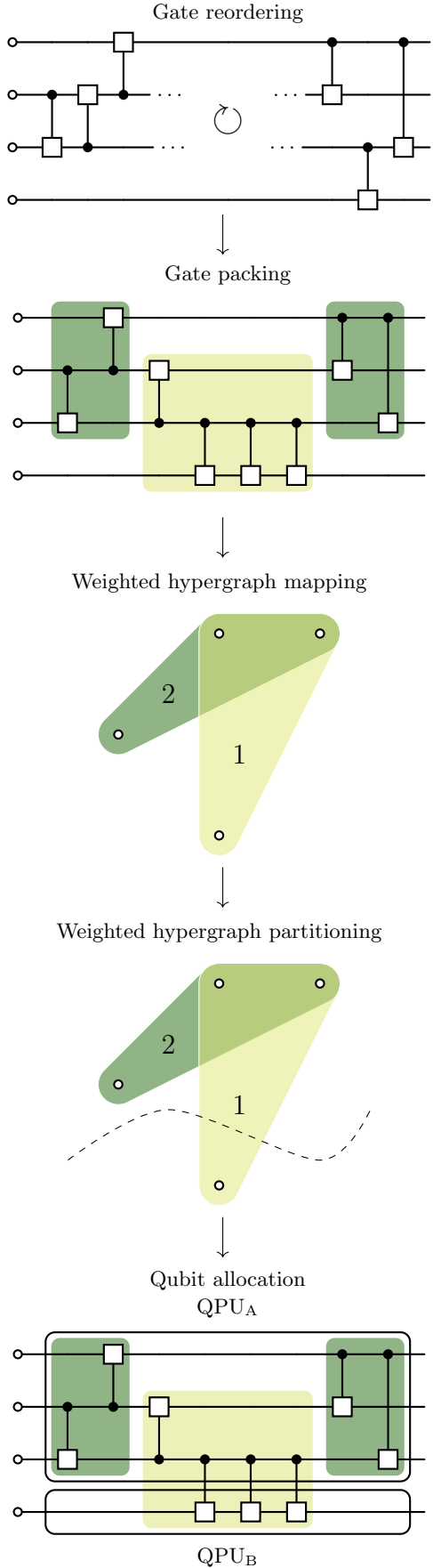

    \centering
    \begin{tikzpicture}
        \node (a) at (0, 0) {
            \import{drawings}{reordering}
        };
        \node (b) at (0, -4) {
            \import{drawings}{packing}
        };
        \node (c) at (0, -9) {
            \begin{tikzpicture}[scale=1.5]
\input{tikz-hypergraph}
\node[circle, draw=black, thick, fill=white, inner sep=0pt, minimum size=3.5pt] (a) at (0, 0) {};  
\node[circle, draw=black, thick, fill=white, inner sep=0pt, minimum size=3.5pt] (b) at (1, 1) {};
\node[circle, draw=black, thick, fill=white, inner sep=0pt, minimum size=3.5pt] (c) at (2, 1) {};
\node[circle, draw=black, thick, fill=white, inner sep=0pt, minimum size=3.5pt] (d) at (1, -1) {};
\node[text=black] (wa) at (0.5, 0.4) {\large{2}};
\node[text=black] (wa) at (1.2, -0.2) {\large{1}};

\begin{pgfonlayer}{bg}
  \draw[draw=white, fill=OliveGreen!50] \hedgeiii{a}{b}{c}{2mm};
  \draw[draw=white, fill=GreenYellow, fill opacity=.5] \hedgeiii{b}{c}{d}{2mm};
\end{pgfonlayer}

\node[draw, color=white, text=black, align=center] at (1, 1.5) {{Weighted hypergraph}
{mapping}};

\end{tikzpicture}
        };
        \node (d) at (0, -14.2) {
            \begin{tikzpicture}[scale=1.5]
\input{tikz-hypergraph}
\node[circle, draw=black, thick, fill=white, inner sep=0pt, minimum size=3.5pt] (a) at (0, 0) {};  
\node[circle, draw=black, thick, fill=white, inner sep=0pt, minimum size=3.5pt] (b) at (1, 1) {};
\node[circle, draw=black, thick, fill=white, inner sep=0pt, minimum size=3.5pt] (c) at (2, 1) {};
\node[circle, draw=black, thick, fill=white, inner sep=0pt, minimum size=3.5pt] (d) at (1, -1) {};
\node[text=black] (wa) at (0.5, 0.4) {\large{2}};
\node[text=black] (wa) at (1.2, -0.2) {\large{1}};

\begin{pgfonlayer}{bg}
  \draw[draw=white, fill=OliveGreen!50] \hedgeiii{a}{b}{c}{2mm};
  \draw[draw=white, fill=GreenYellow, fill opacity=.5] \hedgeiii{b}{c}{d}{2mm};
\end{pgfonlayer}

\draw [black, dashed] plot [smooth, tension=0.5] coordinates { (-0.5, -0.75) (0.5, -0.25) (2, -0.75) (2.5, -0.25)};

\node[draw, color=white, text=black, align=center] at (1, 1.5) {{Weighted hypergraph}
{partitioning}};

\end{tikzpicture}
        };
        \node (e) at (0, -19.5) {
            \import{drawings}{inverse}
        };
        \draw[->] (0., -1.6) to  (0, -2.2);
        \draw[->] (0, -6.1) to  (0, -6.7);
        \draw[->] (0, -11.3) to  (0, -11.9);
        \draw[->] (0, -16.5) to  (0, -17.1);
    \end{tikzpicture}

    \caption{\small Workflow of the araQne compiler: starting from a monolithic circuit, gates are reordered to maximize gate packet size while preserving circuit equivalence (dark and light green boxes in the circuits). The circuit is then transformed into a weighted hypergraph, which is partitioned and mapped to a distributed circuit by allocating qubits to a network of QPUs. This qubit allocation strategy helps reduce the number of required EPR pairs.} 
    \label{fig:Workflow}
\end{figure}

Quantum circuits are sequences of gates defined on a quantum register. Each  qubit is associated to a QPU, and the resulting mapping is called an {\textit{allocation map}}. For a given allocation map, the gates of a distributed circuit are either local (acting on qubits within the same QPU) or non-local (acting on qubits assigned to different QPUs). Implementing non-local gates requires inter-QPU communication. 
Non-local operations are implemented using the TeleGate quantum communication protocol, which consumes one pre-shared EPR pair  for each execution (see Appendix~\ref{app:telegate}).

The overall workflow is presented schematically in Figure \ref{fig:Workflow}. Following the depicted procedure, the compiler generates a distributed version of the input quantum circuit, mapping subcircuits to individual QPUs, while minimizing inter-QPU operations.

In the following, we explain how multiple quantum gates can be distributed together, as a gate packet, using a single TeleGate protocol (Section~\ref{sub:packing-gates}). We present gate reordering strategies which allow us to rewrite the input circuit reducing the distribution cost further (Section~\ref{sub:reordering-gates}). All the mathematical proofs in this section can be found in Appendix~\ref{proofs}. We rely on a hypergraph-based framework to implement the distribution of the circuit, following an approach similar to~\cite{PhysRevA.100.032308} (Section~\ref{sub:mapping-circuits}).

\subsection{Packing quantum gates}
\label{sub:packing-gates}

In order to minimize the distribution cost of a given quantum circuit, one strategy is to distribute multiple quantum gates with the same quantum communication protocol. The notion of gate packet formalizes this strategy, by defining how a sequence of consecutive gates can be distributed together by the same  protocol~\cite{martinez,PhysRevA.100.032308,Burt,Burt2, Wu_2023}. Here, we assume a universal gateset with no specific restrictions on composition, comprising arbitrary single-qubit and two-qubit gates with a matrix representation in
$S=\{CU\mid U\in \mathbb{U}(2) \}\cup \mathbb{SU}(2),$
where $\mathbb{U}(2)$ and $\mathbb{SU}(2)$ are the unitary  and the special unitary groups of $2\times2$ unitary matrices, respectively.

\noindent
\begin{definition}[Gate packet]
Given a circuit $\mathcal C$, a gate packet $\mathcal{P}$ defined on a sub-register $R_\mathcal P$ and rooted on qubit $q\in R_\mathcal P$ is a sequence of consecutive gates such that:
\begin{enumerate}
    \item  each controlled-gate $g\in \mathcal{P}$ is   controlled by $q$,
    \item each single-qubit gate $g \in \mathcal{P}$  acting on $q$ must be either diagonal or anti-diagonal,
    \item for each $t\in R_\mathcal P\setminus \{q\}$ there is at least one controlled-gate $g\in \mathcal P$ such that $g$ targets $t$.
\end{enumerate}
\label{gp}
\end{definition}

\noindent {\textbf{Characterization of a gate packet.}}
The unitary operator associated to a  gate packet defined on a sub-register of size $d$ rooted on $q$ can  be written as \cite{Wu_2023}
\begin{equation}
   \left(\ketbra{0}\cdot D\right) \otimes A+(\ketbra{1} \cdot D) \otimes B  
   \label{gate_packet_identity}
\end{equation}
 where $D\in \{I, X\}$,
 $$
 A=\bigotimes_{\ell=2}^{d}A_\ell \quad\text{ and }\quad B=\bigotimes_{\ell=2}^{d}B_\ell,
 $$
 with $A_\ell, B_\ell\in \mathbb{U}(2)$ for each $\ell\in\{2,\dots, d\}$.
 Equation \eqref{gate_packet_identity} can be visualized as a sequence of controlled gates, each rooted on the same control qubit (here labeled $q_1$) and having different targets as shown in
 Figure~\ref{fig:gate_packet_characterization} {and proved in Appendix~\ref{proofs}. }
 
In order to distribute a circuit $\mathcal C$ to a given network of quantum processors, each qubit in the register has to be assigned to a specific QPU. This process defines an allocation map. 

 \begin{figure}[htpb]
     \centering
    \begin{quantikz}[column sep=0.1cm]
    \phase[style={draw=black, fill=white}]{}&\wire[l][1]["q_1"{above, pos=0.2}]{a} & & \gate{D} & 
  & \ctrl{1} &   & \ctrl{2} & \ \dots\ &  & \ctrl{4}  & \\
    \phase[style={draw=black, fill=white}]{}& \wire[l][1]["q_2"{above, pos=0.2}]{a} & & & \gate{A_2} & \gate{B_2 A_2^\dagger} & & &  \ \dots\ & & & \\
  \phase[style={draw=black, fill=white}]{}& \wire[l][1]["q_3"{above, pos=0.2}]{a} & & & & &  \gate{ A_3}  & \gate{B_3 A_3^\dagger} & \ \dots\ & & & \\
    \wireoverride{n} & \gate[style={draw=white, inner ysep=-2pt}]{\raisebox{5pt}{\vdots}}\wireoverride{n} & \wireoverride{n} & \wireoverride{n} &
    \wireoverride{n} &  \wireoverride{n} &
    \wireoverride{n} \\
  \phase[style={draw=black, fill=white}]{}& \wire[l][1]["q_{d}"{above, pos=0.2}]{a} & & & & & & & \ \dots\ & \gate{ A_{d}} & \gate{B_{d} A_{d}^\dagger} & 
 \end{quantikz}
 
\caption{\small Any gate packet can be expressed as in Equation~\eqref{gate_packet_identity}, and can be implemented using the circuit shown above.}

 \label{fig:gate_packet_characterization}
 \end{figure}

\begin{definition}[Allocation map]
   Let $R$ be the register of qubits in a  quantum circuit and \( L\) a set of labels associated to the available QPUs. An \textit{allocation map}  $\mathcal{A}$ is a map defined as
\begin{equation}
    \mathcal{A}: R \to L
\end{equation}
with $ \mathcal{A}(q_i) = \ell  $ if qubit $q_i$ is assigned to QPU$_\ell$. 
\end{definition}
\noindent

Consider now a gate packet $\mathcal P$ acting on  the sub-register $R_\mathcal P$. The subset of QPUs to which the register $R_\mathcal P$ is mapped by the allocation map $\mathcal A$ can be expressed as 
$\mathcal A(R_\mathcal P)=\{\mathcal A (q):q\in R_\mathcal P\}.$
Equation~\eqref{gate_packet_identity} implies that the distribution cost of a gate packet depends solely on \( |\mathcal{A}(R_\mathcal{P})| \), and is independent of both the number of gates within the packet and the specific qubit allocation. 

\begin{proposition}[Cost of distributing a gate packet]
\label{packet-cost}
Let $\mathcal A$ be any allocation map. Let $\mathcal P$ be any gate packet defined on a sub-register $R_\mathcal P$ of size $d$ such that $|\mathcal A(R_\mathcal P)|=k$. Then, all gates in $\mathcal P$ can be implemented using not less than $k-1$ TeleGate protocols, hence consuming $k-1$  EPR pairs.
\end{proposition}

From there, one can easily compute the distribution cost of a circuit $\mathcal C$.

\begin{definition}[Packing sequence]
A \textit{packing sequence} is a sequence of gate packets 
${\Omega}_\mathcal{C}=  ( \mathcal{P}_1,\dots, \mathcal{P}_m )$
where every $g\in \mathcal C$ is associated to a  unique $\mathcal P\in \Omega_\mathcal{C}$ and such that $\mathcal C$ can be obtained from  $\Omega_\mathcal{C}$ only by adding single qubit gates between its elements.
\end{definition}

\begin{definition}[Distribution cost] 
Given a quantum circuit $\mathcal C$, a $\mathcal C$ packing sequence $\Omega_\mathcal C$ and an allocation map $\mathcal A$,  the distribution cost of $\mathcal C$ is equal to 
\begin{equation}
   \mathcal D(\Omega_\mathcal C)= \sum_{\mathcal P\in \Omega_\mathcal{C}}\left[|\mathcal A(R_\mathcal P)|-1\right].  
    \label{distribution-cost}
\end{equation}
\end{definition}

Consider a  quantum circuit  as a  sequence $\mathcal C = ( g_1, \dots, g_m)$ of $m$ gates acting on a $n$ qubit register $ R= \{q_1, \dots q_n\}$, with $U_\mathcal{C}=g_m \cdots g_1 \label{circuit_op}$ being the unitary operation associated to the circuit $\mathcal C$. 

Given that two circuits are said to be equivalent if they correspond to the same unitary transformation, it is worth noticing that equivalent circuits can lead to different packing sequences.  Hence,  circuit rewriting techniques can help turning the input circuit into an equivalent one that has a different packing sequence, with a lower distribution cost. 
In other words, one can design circuit optimization strategies for DQC. We develop such a strategy in the following section.

\subsection{A greedy {heuristic} algorithm for gate reordering and packing}
\label{sub:reordering-gates}


Note that two adjacent packets $\mathcal{P}_i$ and $\mathcal{P}_{i+1}$ 
can be merged  into a single packet $\mathcal{Q}_j=\mathcal{P}_i \cdot \mathcal{P}_{i+1}$ if they are rooted on the same control qubit. We can now quantify how merging reduces the distribution cost.

\begin{proposition}[Distribution cost after merging]
Given a quantum circuit $\mathcal C$ and its associated packet sequence  $\Omega=(
\mathcal{P}_1,\dots,\mathcal{P}_m )$, if $\Gamma=( \mathcal{Q}_1,\dots,\mathcal{Q}_{\ell} )$  with  $\ell < m$ is a packing sequence constructed from $\Omega$ by merging packets then 
$$
\mathcal D (\Gamma)\leq \mathcal D (\Omega)
$$
for every possible qubit allocation map $\mathcal{A}$.
\end{proposition}

Before partitioning the input quantum circuit, it is crucial to minimize the distribution cost by merging together as many gate packets as possible and distribute them via a single TeleGate protocol. This is achieved with a greedy  approach (see Algorithm~\ref{circuit-reorder}), which makes use of the following choice properties:

\begin{itemize}
    \item {\it{Commutativity of gates}}. Given a circuit $\mathcal C$, if  two adjacent gates $g_i,g_{i+1}$ commute, then the unitary $U_\mathcal C$  is invariant under permutation of $g_i$ and $g_{i+1}$. Algorithm~\ref{circuit-reorder} commutes adjacent gates whenever it helps increasing the size of gate packets, while leaving the unitary operator $U_\mathcal C$ unchanged. 
    \item {\it{Symmetry of controlled-gate}}. A controlled-$U$ gate $CU$ is control-symmetric if it satisfies:
    \begin{align*}
        CU &\equiv \ketbra{ 0}{0} \otimes I + \ketbra{1}{1} \otimes U\\
        &=I\otimes \ketbra{0}{0} +U\otimes \ketbra{1}{1}. 
    \end{align*}
A ${C}U$ gate can be shown to be control-symmetric if and only if $U$ is a phase gate $R_Z(\phi)$. Switching the control qubit of a gate packet can potentially increase the size of a gate packet.
    \item {\it{Packet merging}}. Consecutive gate packets can be merged together to form a larger gate packet, as long as they are rooted in the same control qubit. 
\end{itemize}

\begin{figure}[htpb]
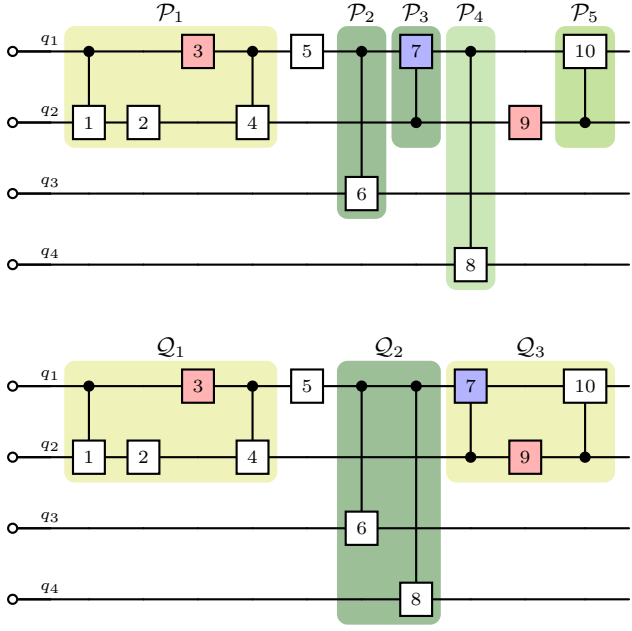

\begin{center}
 \import{drawings}{dist_pack.tex} 
\import{drawings}{reordered-circuit.tex}
\end{center}
\caption{\small (\textit{Top}) Example of gate packets, identified by boxes with shades of green. 
The gate $(3)$ highlighted in red is either  diagonal or anti-diagonal so that $\mathcal{P}_1$ fulfills the definition of a  gate packet.
The numbers are associated with the position of gates as they appear in the figure.
(\textit{Bottom}) Resulting circuit after performing gate-reordering and packet-merging. The gate (7) in blue is assumed to be diagonal and in that case, gates (7) and (8) commute. The packets $\mathcal{Q}_2$ and $\mathcal{Q}_2$ have been obtained by permuting gates (8) and (7) and by merging smaller packets (see text for details). Specifically, packets $\mathcal{P}_2$ and $\mathcal{P}_4$ are merged into packet $\mathcal{Q}_2$, and packets $\mathcal{P}_3$ and $\mathcal{P}_5$ are merged into packet $\mathcal{Q}_2$.}

\label{comm-gates}
\end{figure}

We illustrate this process in Figure~\ref{comm-gates}, where the properties of our greedy approach lead to an equivalent circuit, but with a lower distribution cost. Specifically, suppose that the unitary acting on the target qubit of the control gate (7) (highlighted in blue) is diagonal. In that case, gates (7) and (8) commute and consequently they can be permuted (since the resulting circuit, after permutation, is equivalent to the initial one). After permutation of the gates,  gate (8) can  be added to the packet $\mathcal{P}_2$ to form a larger packet $\mathcal{Q}_2$. Assuming further that gate (9) (highlighted in red in Figure~\ref{comm-gates}) is either diagonal or anti-diagonal, the packets $\mathcal{P}_3$ and $\mathcal{P}_5$   can be merged to create the larger packet $\mathcal{Q}_2$.

\begin{algorithm}[htpb]

\KwData{circuit $\mathcal C$}
\KwResult{packing sequence $\Omega_\mathcal C$}
 
$\Omega_\mathcal C \gets (\,)$\tcp*[r]{Empty sequence}
$A \gets\emptyset$\;
\While{$|A|\neq|\mathcal C|$}{
    $g \gets$ first gate in $\mathcal C$ which is not in $A$\;
    \If{\upshape $g$ is a controlled-gate}{
        $\mathcal P\gets (g)$\tcp*[r]{One element sequence}
        \ForEach{\upshape $\bar g\in\mathcal C$ after $\mathcal P$}{
                \If{\upshape $\bar g\text{ can be merged with }\mathcal P$ \textbf{and} $\bar g$ or $\mathcal P$ commutes with the gates between $\mathcal P$ and $\bar g$}{
                    rearrange $\mathcal C$ to place $\bar g$ and $\mathcal P$ next to each other\;
                    $\mathcal P \gets \mathcal P + (\bar g)$\tcp*[r]{Concatenation}
                }
            } 
            $\Omega_\mathcal{C}\gets \Omega_\mathcal{C}+(\mathcal P)$\;
            $A\gets A\cup\{x\mid x\in \mathcal P\}$\;
        }
        $A\gets A\cup\{g\}$\;
    } 
\Return $\Omega_\mathcal C$\;
\
\caption{Greedy gate reordering and packing algorithm. This algorithm creates larger gate packets using  control-symmetry, commutation and packet merging properties, hence reducing the distribution cost.}
\label{circuit-reorder}
\end{algorithm}

In the remainder of this section, we provide further arguments justifying the relevance of our approach. 
For a given circuit $\mathcal C$, there is a gate packing sequence $\Omega_{min}$ which minimizes the distribution cost of the circuit.
However, in the class of circuits equivalent to $\mathcal C$, there may be circuits that have gate packing sequences with distribution cost lower than $\mathcal D(\Omega_{min})$. One can naturally wonder then whether pairwise commutations of consecutive gates are sufficient to minimize the distribution cost among equivalent circuits.

The answer is affirmative for depth-invariant equivalent circuits. Consider the equational theory of quantum circuits developed in~\cite{complete-eqs}, which establishes that two circuits are equivalent if and only if they can be transformed into one another under a set of elementary circuit equivalences. 

All but two of those circuit equivalences \cite[equalities $(n)$ and $(o)$ in Figure~3]{complete-eqs} are equivalences between circuits of different depth or can be obtained with pairwise permutations of consecutive gates. 

In these two equalities, both sides are circuits with the same constraints on distributability, that is, control and target qubits have identical positions in both circuits. Therefore, applying them does not affect distributability.
 In conclusion, if an equivalent circuit has a gate packing sequence with a lower distribution cost than the minimal distribution cost of the input circuit (for the same depth), it can be obtained through pairwise commutations of consecutive gates. 

\subsection{Weighted hypergraph mapping}
\label{sub:mapping-circuits}

Having explained how gates can be grouped together to lower the distribution cost, we now solve the circuit partitioning problem (i.e. the problem of partitioning a circuit in sub-circuits) to be executed on different QPUs. 
We do so by reducing it to the hypergraph partitioning problem (see Figure \ref{fig:Workflow}).

The input circuit is mapped to a hypergraph, where edges  are gate packets and nodes are qubits. The hypergraph is partitioned to $k$-subgraphs that are then mapped to $k$-subcircuits, in the circuit representation. This partitioning scheme minimizes the number of cut-edges in the hypergraph, corresponding to the minimization of non-local operations between the subcircuits. 

Given a packing sequence $\Omega_\mathcal C$, we map the set of qubits $R=\{q_1, \dots, q_n\}$ to the set of nodes $V$, and  the set of packets sub-registers $\{R_\mathcal P\mid \mathcal P \in \Omega_\mathcal C\}$ to the set of hyperedges $E$. An edge weighting map $w:e\mapsto w_e$ is introduced for each $e\in E$ where
$w_e=|\{\mathcal P\in \Omega_\mathcal C\mid R_\mathcal P=e\}|$ identifies the number of times that
the packet subregister $e$ appears in $\Omega_\mathcal{C}$.

A $k$-partitioning of the circuit which minimizes the distribution cost corresponds then to the $k$-partitioning of the hypergraph $H=(V, E, w)$. Indeed, the objective of the hypergraph $k$-partitioning is to find a partition $\Pi=\{V_i\}_i$, with $i\in \{1, \dots, k\}$, of $V$ so as to minimize the cost defined as the sum:
$$
\sum_{e\in E}[|F(e)|-1]w(e)
$$
where $F(e)=\{X\in \Pi\mid X\cap e\neq\emptyset\}$ is the set of parts which contain elements to which $e$ is adjacent. 

\begin{figure}[h]
    \centering
\begin{tikzpicture}[scale=1.5]
\input{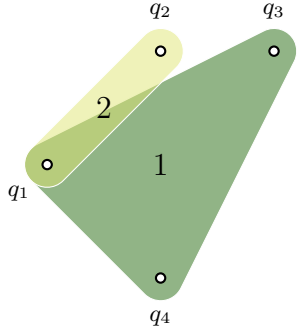}
\node[circle, draw=black, thick, fill=white, inner sep=0pt, minimum size=3.5pt] (a) at (0, 0) {};  
\node[circle, draw=black, thick, fill=white, inner sep=0pt, minimum size=3.5pt] (b) at (1, 1) {};
\node[circle, draw=black, thick, fill=white, inner sep=0pt, minimum size=3.5pt, text=black] (c) at (2, 1) {};
\node[circle, draw=black, thick, fill=white, inner sep=0pt, minimum size=3.5pt] (d) at (1, -1) {};
\node[circle, inner sep=0] (wa) at (0.5, 0.5) {\large{2}};
\node[circle, draw=OliveGreen!50, fill=OliveGreen!50, text=black, inner sep=0] (wa) at (1, 0) {\large{1}};

\node[draw, color=white, text=black, draw opacity=0] at (2, 1.35) {$q_3$};
\node[draw, color=white, text=black, draw opacity=0] at (1, 1.35) {$q_2$};
\node[draw, color=white, text=black, draw opacity=0] at (1, -1.35) {$q_4$};
\node[draw, color=white, text=black, draw opacity=0] at (-0.25, -0.25) {$q_1$};

\begin{pgfonlayer}{bg}
  \draw[draw=white, fill=OliveGreen!50] \hedgeiii{d}{a}{c}{2mm};
  \draw[draw=white, fill=GreenYellow, fill opacity=.5] \hedgeii{a}{b}{2mm};
\end{pgfonlayer}

\end{tikzpicture}

    \makeatletter
        \long\def\@ifdim#1#2#3{#2}
    \makeatother
    \caption{\small Weighted hypergraph obtained from the circuit in Fig.~\ref{comm-gates}. The dark green hyperedge corresponds to packet $\mathcal{Q}_2$, which is the only packet acting on qubits $\{q_1, q_3, q_4\}$, and therefore has weight $1$. In contrast, the light green hyperedge has weight $2$ because two packets, namely $\mathcal{P}_1$ and $\mathcal{Q}_3$, act on qubits $\{q_1, q_2\}$.}
    
    \label{example-HG-Map}
\end{figure}

Since an allocation map $\mathcal A$ corresponds to a partition $\Pi$, one can see that the distribution cost is equal to the cost minimized by the hypergraph $k$-partitioning:
$$
\begin{aligned}
        \mathcal D(\Omega_\mathcal C)&=\sum_{e\in E}\sum_{\{\mathcal P\in\Omega_\mathcal C\mid R_\mathcal P=e\}} [|\mathcal A(e)|-1]\\
    &=\sum_{e\in E} [|\mathcal A(e)|-1] w_e.
\end{aligned}
$$
A pseudo-code that implements the explained mapping of a circuit into a weighted hypergraph is provided in Algorithm~\ref{circuit-to-hypergraph}. Applying this algorithm to the circuit in Figure~\ref{comm-gates}, we obtain the weighted hypergraph shown in Figure~\ref{example-HG-Map}.
\begin{algorithm}[h]
\KwData{packing sequence $\Omega_\mathcal C$}
\KwResult{weighted hypergraph $H=(V, E, w)$}
$V \gets \emptyset$\;
$E\gets \emptyset$\;
$w \gets \emptyset$\;

\ForEach{\upshape $\mathcal{P}_i\in \Omega_\mathcal C$}{
    $V \gets V \cup R_i$\;
    $E \gets E \cup \{R_i\}$ \;
}

\ForEach{\upshape $e\in E$}{
    $w_e \gets |\{\mathcal P_i\in \Omega_\mathcal C\mid R_i=e \}|$\;
    $w \gets w \cup \{w_e\}$\;
}
\Return $H=(V, E,w)$\
\vspace{10pt}
\caption{Weighted hypergraph construction. This algorithm constructs a weighted hypergraph from a sequence of packets.}
\label{circuit-to-hypergraph}
\end{algorithm}

 Assessing the ability of our algorithm to perform near-optimal choices requires to benchmark the quality of its results on standard quantum algorithms, as realized in the next section.

\section{Distribution cost benchmarking}
\label{sec:benchmarks}

We benchmark the performance of our  algorithm on several classes of quantum circuits. We suppose that the set of $k$ QPUs forming the quantum network is made of identical QPUs with an all-to-all connectivity and where each QPU is interconnected to all other QPUs. Given a monolithic circuit of $n$ qubits, a balanced problem is considered, where each quantum processor approximately contains $\frac{n}{k}$ qubits. To assess the performance of our approach, we first consider the case of random-generated quantum circuits. Then, we provide performance results on circuits from the QASMBench 1.4 benchmark, a widely used benchmarking suite for quantum circuits~\cite{qasm}.

\subsection{Random Quantum Circuits}

We consider here the case of $n$-qubit circuits constructed as sequences of gates randomly selected from the sets of gates $G_1$, $G_2$ defined as:
$$
G_1=\{H, X, RZ(\theta)\}\quad\text{ and } \quad G_2=\{CX, CRZ(\theta)\}
$$ 
where $\theta\in[0, 2\pi]$.
This choice of gates for $G_1, G_2$ is motivated by the fact that (anti-)diagonal and non-diagonal single-qubit gates, as well as control-symmetric and non-control-symmetric gates are included in them. As explained in Section~\ref{sec:workflow}, (anti-)diagonal and control-symmetric gates play an important role in the
construction of gate packets. As a consequence, including them in the gateset to generate random circuits provides better insights into the performance of our algorithm.

The construction of a random circuit proceeds as follows: we first randomly draw an angle $\theta$ and two qubits from a uniform probability distribution. In a next step, a gate $g$ is randomly selected to be a one- or two-qubit gate with a probability $p = 0.2$ and $1-p$, respectively 
\footnote{Note that there is nothing particular about this value of $p$. The only criterion for fixing $p$ is that the random circuit should include a large proportion
of two-qubit gates.}.

In both cases, the selection of gates within a given set, namely $G_1$ or $G_2$, is done uniformly at random. More explicitly, if $g$ is a one-qubit gate, it
is chosen among the gates in $G_1$ with equal probability for all gates.
By design, if $g$ belongs to $G_1$, two gates of type $g$ will both act on previously selected qubits.
When $g$ is a two-qubit-gate in $G_2$, one of the two qubits it acts on is chosen as the control qubit uniformly at random. 
By construction, the process continues until $n^2$ gates from $G_2$ have been added to the circuit.

\begin{figure}[h]
\begin{center}
\includegraphics[width=280pt]{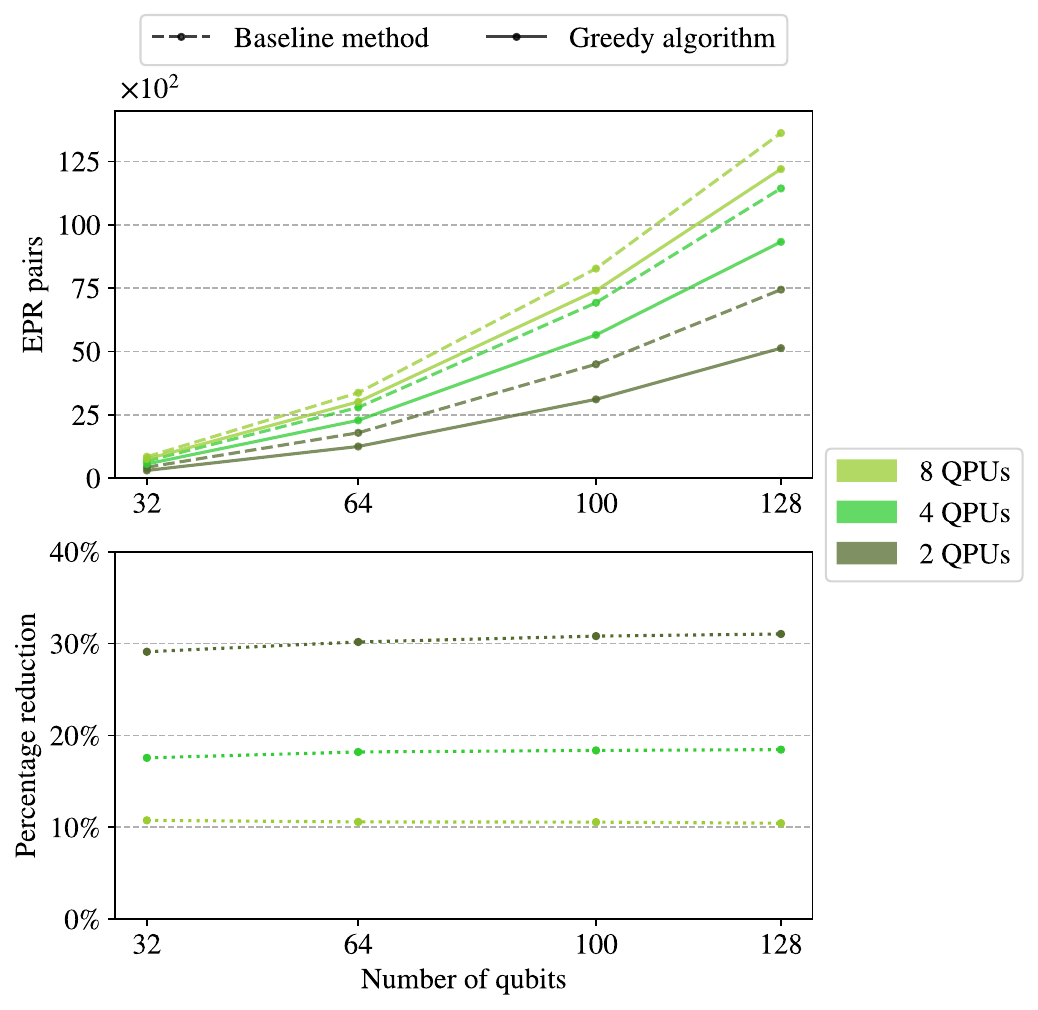}
\end{center}
\caption{\small 
(\textit{Top}) Distribution cost (as the number of EPR pairs consumed) for the partitioning of input random circuits into 2,4, and 8 QPUs. Results are
displayed, as a function of the overall number of qubits in the circuit, using (i) our approach based on gate-reordering (solid lines) and (ii) baseline method where gates are grouped without considering possible reordering (dashed lines). (\textit{Bottom}) Reduction (expressed as a percentage) in the number of EPR pairs obtained when using the greedy algorithm with gate reodering instead of the baseline method.
}
\label{random_fig}
\end{figure}

We compare results obtained with our greedy approach based on gate-reordering and a baseline method that groups consecutive controlled gates without applying any reordering. Results for the number of EPR pairs required for the partition of input circuits into 2, 4 and 8 QPUs are shown in Figure~\ref{random_fig} as a function of  number of qubits per QPU. To avoid bias arising from the randomness in circuit construction, we average our results over 10 independent repetitions, i.e., 10 randomly generated circuits per data point.

As one can see in Figure~\ref{random_fig}, our approach clearly allows for a reduction in  the number of EPR pairs in comparison to the baseline method. We show also in the lower part of Figure~\ref{random_fig} the reduction, expressed as a percentage, in the number of EPR pairs obtained using the greedy approach instead of the baseline method. The results show that for a partition into 2 QPUs 
 the average reduction is approximately $30\%$, for 4 QPUs, the improvement is around $19\%$, while for 8 QPUs, one gains a $10\%$ reduction factor. 

From the above results, the EPR pair reduction ratio obtained with our method  decreases for an increasing number of QPUs. This feature can be traced back to the heuristic nature of the presented algorithm (\ref{circuit-reorder}), and to the fact that the complexity of the DQC problem increases with the number of QPUs. Finally, note that in the limit of a large number of QPUs, the distribution cost is bounded by the number of control gates in the circuit, which in this case was set to $n^2$.

\subsection{QASMBench quantum circuit benchmarks}
We now benchmark our approach with quantum circuits from the QASMBench benchmark suite \cite{qasm}, in its version 1.4. Table~\ref{tab_qasm} aggregates the results for the number of EPR pairs obtained for a partition into 2 QPUs using, as in the previous case of random circuits, (i) the greedy algorithm and (ii) a baseline method.
For all cases considered here, computations are performed with the transpiled version of the OpenQASM circuit \cite{qasm}.

\begin{table}[h]
\centering
\begin{tabular}{lccc}
\hline
\textbf{Circuit (qubits)}  & \quad\makecell{\textbf{Greedy}\\\textbf{algorithm}} \qquad& \makecell{\textbf{Baseline}\\\textbf{method}} \\
\hline
Adder (28)         & 7    & 9    \\
Adder (64, 118)     & 7    & 10   \\
BV (70, 140)       & 1    & 1    \\
Cat (35)           & 1    & 1    \\
DNN (33)           & 19   & 27   \\
DNN (51)           & 29   & 42   \\
GHZ (40)           & 1    & 1    \\
Ising (34)         & 1    & 1    \\
KNN (41)           & 1    & 1    \\
Multiplier (45)    & 162  & 238  \\
Multiplier (75)    & 380  & 568  \\
Multiplier (350)   & 6397  & 10954  \\
Multiplier (400)   & 8319  & 13492  \\
QFT (29)           & 14   & 210  \\ 
QFT (63)           & 36   & 739  \\
QFT (160, 320)      & 40   & 820  \\
QRAM (20)          & 18   & 33   \\
QuGAN (39)         & 24   & 34   \\
QuGAN (71)         & 38   & 55   \\
QuGAN (111)        & 58   & 85   \\
QuGAN (395)        & 196  & 292  \\
QV (32)            & 600  & 600  \\
QV (100)           & 6492 & 6492 \\
Swap Test (25)     & 12   & 18   \\
Swap Test (41)     & 20   & 30   \\
Swap Test (83)     & 40   & 60   \\
W-State (76, 118)   & 2    & 2    \\
Square root (45)   & 3418 & 4419 \\
\hline
\end{tabular}
\caption{\small Distribution cost (measured as the number of EPR pairs consumed) for several circuits in the QASMBench 1.4 benchmarking suite~\cite{qasm}. For each circuit, its name and number of qubits 
$n$ are specified. We partition all the circuits across two QPUs, each containing 
$\lfloor n/2 \rfloor +1$ qubits, without making any assumptions about the intra-QPU topology. Results are shown using the greedy algorithm and a baseline implementation with no reordering. For all circuits considered here, the transpiled version was used \cite{qasm}.}
\label{tab_qasm}
\end{table}

As seen in Table~\ref{tab_qasm}, the greedy algorithm allows for an overall significant reduction in the number of EPR pairs
with respect to the baseline method. There are few noticeable exceptions to this pattern such as for circuits where the minimum number of EPR pairs required for inter-QPU communication, which is equal to $1$ in this bipartition case, is already reached by the baseline method (see results for the Ising and KNN circuits in Table~\ref{tab_qasm}). Other noticeable cases include the Quantum Volume circuits \cite{cross}, namely QV32, QV100 where the greedy algorithm does not enable any reduction in the number of EPR pairs. This is naturally explained by the fact that for these circuits,
commutation between gates is not possible and consequently our gate reordering procedure has no impact on results.  

Table~\ref{tab_qasm} includes several results for Quantum Fourier Transform (QFT) circuits with a qubit number ranging from 29 to 320. 
The values of rotation angles in QFT circuits can be expressed as $2^{-k}$ with $k \leq n$ and $n$ the total number of qubits. Naturally, as the qubit number $n$ increases, there are more rotation matrices with a small enough angle that can be replaced, without impacting results, by the Identity matrix \footnote{our compiler uses some subroutines from Qiskit and in this context, a rotation matrix with an angle smaller than $10^{-10}$ is replaced by the identity matrix.}. In other words, as $n$ increases, larger subparts
of the circuit are not affected by the algorithm and consequently will not have an impact on the overall distribution cost. This results in a stagnation of the number of EPR pairs computed with the greedy algorithm at $40$ (see results in Table~\ref{tab_qasm} for QFT (160, 320)) for partitioning the larger QFT circuits.

Among the recent new approaches dedicated to DQC, the work in \cite{kaur2025optimizedquantumcircuitpartitioning} allows for a direct comparison
with our results  on the number of EPR pairs obtained with the greedy \ algorithm. The algorithm for distributing a quantum circuit in \cite{kaur2025optimizedquantumcircuitpartitioning} is based on a graph description of circuits where both qubit and gate teleportations are considered. 
By comparing the best results reported in Table 1 in \cite{kaur2025optimizedquantumcircuitpartitioning}, 
we can see that in most cases, our greedy approach allows to reach a lower (or equal) number of EPR pairs for a bipartition of QASM circuits. Only 
for the circuits Multiplier (350, 400) and the Quantum Volume QV (100), the greedy  algorithm results in higher values for the number of EPR pairs. We want to point out, as already mentioned in previous works such as \cite{Burt}, that gate teleportation is not the most efficient communication protocol to distribute a QV circuit and that in this context, qubit teleportation is more suited for partitioning. This clearly explains why, for the QV circuits, we obtain larger values for the number of EPR pairs  than  the approach in \cite{kaur2025optimizedquantumcircuitpartitioning}.

\vspace{10pt}
\section{Conclusion}

In this work, 
we have presented the greedy heuristic algorithm for efficient gate reordering and packing and the partitioning approach embedded in \textit{araQne}, the distributed quantum compiler developed by Welinq. 

For an input monolithic quantum circuit, the algorithm reorders the gates and groups them into gate packets. Since all gates in a packet   can be distributed using a single TeleGate protocol, this method leads to a lower consumption of entanglement resources. The resulting quantum circuit, expressed as a sequence of gate packets, is then mapped to a weighted hypergraph, whose hyperedges and vertices are associated with gate packets and qubits, respectively. In this way, the circuit partitioning problem is reduced to a hypergraph partitioning problem. 

We showed that the proposed greedy algorithm yields a significant reduction in distribution cost for random quantum circuits compared to a baseline approach where no reordering is applied. We also benchmarked our compiler on the QASMBench 1.4 benchmarking suite. Our numerical results establish the crucial role that circuit reordering strategies play in reducing the cost of distributing quantum circuits onto interconnected QPUs.

 We are planning to integrate additional features to the compiler algorithm in order to enhance further its current performance and introduce hardware awareness. A wider set of communication protocols, such as TeleData and multipartite entanglement-based protocols \cite{Riera_S_bat_2024}, will be added, for a further reduction of the distribution cost obtained with TeleGate alone. In fact, depending on the compiled circuit, a choice of the use of a single or multiple protocols can be favored to minimize the distribution cost. In addition, we will consider techniques that generalize the notion of gate packets, such as embedding ~\cite{martinez,Wu_2023}.
Finally, we will introduce techniques that take into account the connectivity constraints of a given architecture, both at a global level (inter-QPU) and at a local level (coupling map), combined with the fidelity of quantum links in the distribution cost.

\begin{acknowledgments}
This research has received funding from the Région Ile-de-France as part of the project AQADOC (Convention No.~24003067), from the European Innovation Council (EIC) Accelerator under Grant Agreement No.~101188682 for the project SQOUT and from the project Hybrid HPC Quantum Initiative (HQI) which is part of Plan France 2030. J.L. is a member of the Institut Universitaire de France.
\end{acknowledgments}

\bibliography{refs}

\renewcommand{\theproposition}{B.\arabic{proposition}}
\setcounter{proposition}{0}

\appendix
\section{Gate teleportation}
\label{app:telegate}

\begin{figure}[h]
\begin{center}
 \import{drawings}{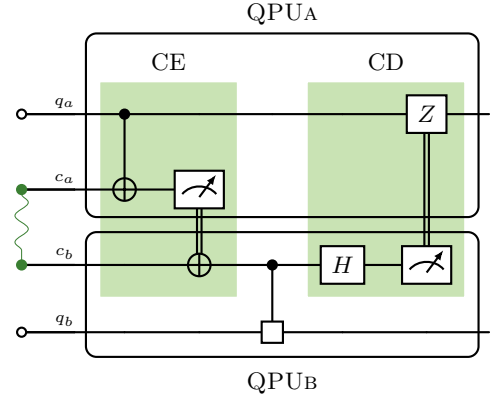}
\end{center}
\caption{\small TeleGate protocol implementing a non-local gate between the data qubits $q_a$ and $q_b$, which belong to distinct QPUs (A and B). The protocol incorporates the Cat-Entangler (CE) primitive, allowing for the second half of the entangled pair to act as a shared copy of the original control qubit. Finally, the shared state of the control qubit $q_a$ is restored at $\text{QPU}_\text{A}$ via the Cat-Disentangler (CD) primitive.}
\label{telegate-figure}
\end{figure}

The TeleGate protocol enables the implementation of a two qubit gate involving qubits which belong to separate QPUs. As seen in Figure~\ref{telegate-figure}, the Cat-Entangler (CE) allows for the second half of the entangled pair to act as a shared copy of the original control qubit. We justify the use of the TeleGate protocol in this section.

To illustrate how to perform inter-QPU quantum operations, assume that there are two qubits $q_a$ and $q_b$ which belong to the two distinct QPUs, respectively QPU$_\text{A}$ and QPU$_\text{B}$. If a non-local controlled-gate with control qubit $q_a$ and target qubit $q_b$ has to be performed, then two kind of protocols can be applied: TeleData or TeleGate.
The starting point of both protocols is the efficient storing of an EPR pair into two communication qubits  each belonging to a distinct QPU.
 
In the case of the TeleData protocol, quantum teleportation transfers a qubit’s state between QPUs to enable local execution of non-local operations \cite{sundaram2023distributingquantumcircuitsusing}. As the data qubit from the first QPU is measured and discarded, restoring the original allocation requires a second teleportation.

On the other hand, the TeleGate protocol leverages two key communication primitives, the Cat-Entangler (CE) and Cat-Disentangler (CD) \cite{Eisert_2000} to execute a non-local controlled-gate. Figure~\ref{telegate-figure} illustrates one method of implementing a non-local  operation using the CE and CD primitives. 
Unlike the TeleData protocol, only the communication qubits are measured, while the data qubits remain in their original partitions.
This implies that the qubit allocation remains unchanged after applying the TeleGate protocol. Consequently, the distribution cost associated with TeleGate corresponds to the consumption of a single EPR pair. Our choice of TeleGate as a quantum communication protocol is justified further by the property that more than one gate can be distributed with a single TeleGate~\cite{yimsiriwattana2004generalizedghzstatesdistributed}.

\section{Detailed proofs}
\label{proofs}
In the following, the proofs of the propositions stated in the main text are presented. These concern the distribution cost of a packet and the resulting cost after merging.

\subsection{Cost of distributing a packet}

Define the projector \( P_i = \ketbra{i} \), and the \( d \)-fold identity operator as $I_d = \bigotimes_{\ell=1}^d I$.
Additionally, let \( R_T \subset \{q_1, \dots, q_n\} \) be a sub-register such that \( |R_T| < n \) and \( q_i \notin R_T \) and \( U \in \mathbb{U}(2^n) \) be a unitary operator. We define the controlled gate \( CU(R_T) \), with control qubit \( q_i \) and target qubits in \( R_T \), as the operation that applies \( U \) on \( R_T \) conditioned on the state of \( q_i \). For simplicity, we denote this operation as \( CU \) when the control qubit and the target sub-register are clear from the context.

\begin{proposition}[Characterization of a gate packet]
The unitary operator implemented by all the gates contained in any gate packet defined on a sub-register of size $d$ rooted on $q$ can always be written in the following form:
$$
    (\ketbra{0}{0}\cdot D) \otimes A+(\ketbra{1}{1}\cdot D) \otimes B
$$
where $D\in \{I, X\}$, 
$A=\bigotimes_{\ell=2}^{d}A_\ell,
B=\bigotimes_{\ell=2}^{d}B_\ell,$
with $A_\ell, B_\ell\in \mathbb{U}(2)$ for each $\ell\in\{2,\dots, d\}$.
\label{prop_0}
\end{proposition}

\begin{proof}
Assume that without loss of generality, we have $R_\mathcal P=\{q_1, \dots, q_{d}\}$ and $q=q_1$ (up to a permutation of qubit labels). Now, let us write $R_T=\{q_2, \dots, q_{d}\}$, and observe the following:
 \begin{itemize}
     \item any product of diagonal and anti-diagonal operators in $\mathbb{U}(2)$ is equivalent to a diagonal or anti-diagonal operator if the number of anti-diagonal operators in the product is even or odd, respectively;
    \item any product of single-qubit operators acting on the register $R_\mathcal{P}\setminus\{q_0\}$ can always be written as 
    $W=\bigotimes_{\ell=2}^{d} W_\ell\in \mathbb{U}\left(2^d\right)$
    with $W_\ell\in\mathbb{U}(2)$;
   \item any product of $s$ control gates with same control qubit  is a control gate. Indeed 
     \begin{align*}
         \prod_{i=1}^sCU^i(q_i)&=\prod_{i=1}^s\left(P_0\otimes I_n+P_1\otimes U^i\right)\\
         &=P_0\otimes I_n+P_1\otimes \left(\prod_{i=1}^s U^i\right)\\
         &=P_0\otimes I_n+P_1\otimes U=CU\left(R_T\right)
     \end{align*}
     where $U^i\in \mathbb{U}(2^d)$ and
     $$
     U=\prod_{i=1}^s \left(\bigotimes_{\ell=2}^{d}U_\ell^i\right)=\bigotimes_{\ell=2}^{d}\left(\prod_{i=1}^s U_\ell^i\right)=\bigotimes_{\ell=2}^{d} U_\ell
     $$
     with $U_\ell\in\mathbb{U}(2)$, because of the gateset considered.
 \end{itemize} 

This means that, without loss of generality, it is always possible to write any gate packet as follows:
\begin{equation}
   \prod_{i=1}^{h} CU^i\cdot\left(V^i\otimes W^i\right)=\prod_{i=1}^h \mathcal Q_i
   \label{eq:figure}
\end{equation}
where  $V^i\in \mathbb{U}(2)$ is a diagonal or anti-diagonal operator,
\begin{equation}
\label{factorization}
U^i=\bigotimes_{\ell=2}^{d}  U^i_\ell\text{\quad and\quad}  W^i=\bigotimes_{\ell=2}^{d} W^i_\ell
\end{equation}
with $U_\ell^i$ and $W_\ell^i$ in $\mathbb{U}(2)$.

First, it is proven that:
\begin{equation}
    \label{first-property}
    \mathcal Q_i=(P_0\cdot D_i) \otimes L_i+(P_1\cdot D_i) \otimes  B_i
\end{equation}
with $D_i\in \{I, X\}$.  In order to do that, two cases have to be considered.
\begin{enumerate}
    \item $V^i$ is diagonal, that is
    $$
    V^i=\begin{bmatrix}
        L_i & 0 \\
        0 & b_i
    \end{bmatrix}=L_iP_0+b_iP_1.
    $$
    In this case it is possible to write
    \begin{align*}
    \mathcal Q_i&=(P_0\cdot I)\otimes \left(L_i W^i\right) + (P_1\cdot I) \otimes \left(b_iU^i\cdot  W^i\right)\\
    &=(P_0\cdot D_i) \otimes L_i+(P_1\cdot D_i) \otimes B_i
    \end{align*}
    where $L_i=L_i W^i$, $B_i=b_iU^i \cdot W^i$ and $D_i=I$.
    \item $V^i$ is anti-diagonal, i.e. : 
    $$
    V^i=\begin{bmatrix}
        0 & L_i \\
        b_i & 0
    \end{bmatrix}=L_i\ketbra{0}{1}+b_i\ketbra{1}{0}.
    $$
    In this case, noticing that $P_0\cdot X=\ketbra{0}{1}$ and $P_1\cdot X=\ketbra{1}{0}$, it is possible to write
    \begin{align*}
    \mathcal Q_i&=(P_0\cdot X)\otimes \left(L_i W^i\right) + (P_1\cdot X) \otimes \left(b_iU^i\cdot  W^i\right)\\
    &=(P_0\cdot D_i) \otimes L_i+(P_1\cdot D_i) \otimes B_i
    \end{align*}
    where $L_i=L_i W^i$, $B_i=b_iU^i \cdot W^i$ and $D_i=X$.
\end{enumerate}

\begin{figure*}
   \import{drawings}{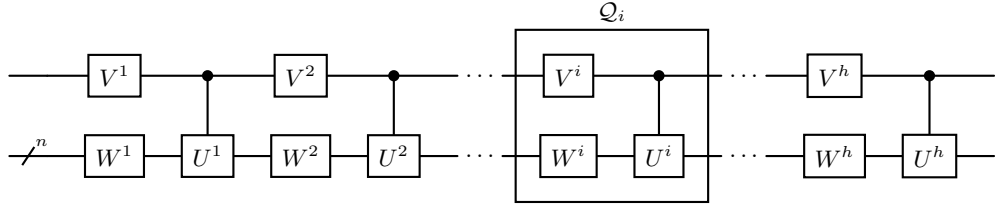}
    \caption{Circuit representation of the operator in equation \eqref{eq:figure}. Every gate packet can be written in the form given by this operator and thus can be represented in this way.}
\end{figure*}

Now we show that
\begin{equation}
     \label{second-property}
    Q_i\cdot \mathcal Q_j=(P_0\cdot D) \otimes \alpha+(P_1\cdot D) \otimes \beta
\end{equation}
by distinguishing the following two cases.

\begin{enumerate}
    \item  $D_i=I$. In this case, it is possible to write the composition of two consecutive operators $Q_i$ and $Q_j$ as
    \begin{align*}
        \mathcal Q_i\cdot \mathcal Q_j &= (P_0\cdot D_j)\otimes \left(L_i\cdot A_j\right)+\left(P_1\cdot D_j\right)\otimes \left(B_i\cdot B_j\right)\\
        &=\left(P_0\cdot D\right)\otimes \alpha+(P_1\cdot D)\otimes \beta
    \end{align*}
    where $C=D_j\in\{I, X\}$, $\alpha=L_i\cdot A_j$ and $\beta=B_i\cdot B_j$.
    \item $D_i=X$. In this case 
    \begin{align*}
        \mathcal Q_i\cdot \mathcal Q_j &= (\ketbra{0}{1}\cdot D_j)\otimes \left(L_i\cdot B_j\right)+\left(\ketbra{1}{0}\cdot D_j\right)\otimes \left(B_i\cdot A_j\right)\\
        &=(P_0\cdot D)\otimes \alpha+(P_1\cdot D)\otimes \beta
    \end{align*}
    where $D=I$ if $D_j=X$ and $D=X$ if $D_j=I$, $\alpha=L_i\cdot B_j$ and $\beta=B_i\cdot A_j$.
\end{enumerate}

Equations \eqref{first-property} and \eqref{second-property} imply that any assignment packet can always be written as:
$$
(P_0\cdot D) \otimes A+(P_1\cdot D) \otimes B
$$
with $C\in\{I, X\}$ and $A, B\in\mathbb{U}(2^d)$.

Finally, the fact that
$$
A=\bigotimes_{\ell=2}^{d}A_\ell \text{\quad and \quad} B=\bigotimes_{\ell=2}^{d}B_\ell, 
$$
with $A_\ell, B_\ell\in \mathbb{U}(2)$, is an immediate consequence of equation \eqref{factorization}.
\end{proof}

From this proposition, the following corollary follows:

\begin{proposition}[Cost of distributing a gate packet]
Let $\mathcal A$ be any allocation map. Let $\mathcal P$ be any gate packet defined on a sub-register $R_\mathcal P$ of size $d$ such that $|\mathcal A(R_\mathcal P)|=k$. Then, all gates in $\mathcal P$ can be implemented using not less than $k-1$ TeleGate protocols, hence consuming $k-1$  EPR pairs.
\end{proposition}

\begin{proof}
We proceed by induction on $k=|\mathcal A(R_\mathcal P)|$.

For $k=1$, no EPR pair is required to implement local gates that act on qubits managed by the same QPU. For $k=2$, the gate packet $\mathcal P$ satisfies the distributability condition introduced in \cite{Wu_2023} and thus it is possible to distribute it using a single EPR pair.

When $k>2$, it is possible to assume that the sub-register $R_a=\{ q_{n_{a-1}},  q_{n_{a-1}+1}, \dots, q_{n_a-1}, q_{n_a}\}$ with $n_{a-1}<n_a$ is allocated to QPU $a$ for each $a\in\{1, \dots, k\}$, up to a permutation of the qubit and QPU indices. Introducing $n_0=0$, noticing that $n_k =d$, and using the  Prop.~\ref{prop_0}, the unitary implementing $\mathcal P$ can be written as
\begin{equation}
(P_0\cdot D) \otimes \left(\bigotimes_{\ell=2}^{d} A_\ell\right)+(P_1\cdot D) \otimes  \left(\bigotimes_{\ell=2}^{d} B_\ell\right)=\left(\prod_{a=1}^k \mathcal P_a\right)\cdot (D\otimes I_n)
\label{second-repr}
\end{equation}
where
\begin{widetext}
\begin{align*}
\mathcal{P}_a&=P_0\otimes I_{n_{a-1}}\otimes \left(\bigotimes_{\ell=n_{a-1}}^{n_a}A_\ell\right)\otimes I_{d-n_a}+P_1\otimes I_{n_{a-1}}\otimes \left(\bigotimes _{\ell=n_{a-1}}^{n_a}B_\ell\right)\otimes I_{d-n_a}\\
&= P_0\otimes I_{n_{a-1}}\otimes \alpha_a\otimes I_{d-n_a}+P_1\otimes I_{n_{a-1}}\otimes \beta_a\otimes I_{d-n_a}.
\end{align*}
\end{widetext}

It is then possible to consider $\mathcal{P}$  as a sequence of $k$ packets rooted on $q_1$ 
$$
\mathcal{P}= (
C\otimes I_n, \mathcal{P}_1, \mathcal{P}_2,\dots,\mathcal{P}_{k} )
$$
where $|\mathcal A(R_{\mathcal P_1})|=1$ and $|\mathcal A(R_{\mathcal 
P_a})|=2$ for each $a\in \{2, \dots, k\}$. Therefore, every $\mathcal P_a$ with $a>2$ can be distributed using a single EPR pair, which implies that $\mathcal{P}$ can be executed using $k-1$ EPR pairs.

\end{proof}

\subsection{Distribution cost after merging}

\begin{proposition}[Distribution cost after merging]
    Given a quantum circuit $\mathcal C$ and its associated packet sequence  $\Omega=(
\mathcal{P}_1,\dots,\mathcal{P}_m )$, if  $\Gamma=( \mathcal{Q}_1,\dots,\mathcal{Q}_{\ell} )$  with  $\ell < m$ is a packing sequences constructed from $\Omega$ by means of merging packets then 
$$
\mathcal D (\Gamma)\leq \mathcal D (\Omega)
$$
for each qubit allocation map $\mathcal{A}$.
\end{proposition}

\begin{proof}
Let $\mathcal{P}_i $ and $ \mathcal{P}_{i+1}$ two adjacent packets that can be merged into a packet
$\mathcal{Q}_j= \mathcal{P}_i \cdot \mathcal{P}_{i+1}$, with $q$ being the  root control qubit. We distinguish the following two cases.
\begin{itemize}
\item $\mathcal{A}(R_{\mathcal{P}_i}) \cap \mathcal{A}\left(R_{\mathcal{P}_{i+1}}\right) \supsetneq \mathcal{A}(q)$. In this case it follows that
$$
\begin{aligned}
    | \mathcal A\left(R_{\mathcal{Q}_j}\right)|&<| \mathcal A\left(R_{\mathcal{Q}_j}\right)|+\left| \mathcal{A}(R_{\mathcal{P}_i}) \cap \mathcal{A}\left(R_{\mathcal{P}_{i+1}}\right)\setminus \mathcal A(q) \right|\\
    &= | \mathcal A\left(R_{\mathcal{P}_i}\right)|+\left|\mathcal A\left(R_{\mathcal{P}_{i+1}}\right)\right|.
\end{aligned}
$$
Indeed, only one TeleGate is needed to perform a non-local operation involving a common allocated QPU different from QPU$_{\mathcal A(q)}$, instead of two. Hence in this case $D(\Gamma)<D(\Omega)$.
\item $\mathcal{A}(R_{\mathcal{P}_i}) \cap \mathcal{A}\left(R_{\mathcal{P}_{i+1}}\right) = \mathcal{A}(q)$. In this case it follows that 
$$
\begin{aligned}
     &| \mathcal A\left(R_{\mathcal{Q}_j}\right)|=
     |\mathcal A\left(R_{\mathcal{P}_i}\right)|+\left|\mathcal A\left(R_{\mathcal{P}_{i+1}}\right)\right|.
\end{aligned}
$$ 
Hence there is no advantage in merging because $D(\Gamma)=D(\Omega)$.
\end{itemize}
Since $\Gamma$ is a packing sequence constructed from $\Omega$ only by merging packets then the proposition follows.
\end{proof}

\end{document}